\documentclass[11pt,onecolumn,twoside]{IEEEtran}

\usepackage{amsfonts,amsmath, amssymb, graphicx, cite}

\newtheorem{defin}{Definition}

\begin{document}

\title{A Big Data Approach to Computational Creativity%
\thanks{
  This work is based in part on a paper presented at the IEEE International Conference on Cognitive Informatics
  and Cognitive Computing, New York, July 2013 \cite{VarshneyPVSC2013}, an abstract presented at the Workshop on 
  Information in Networks, New York, October 2013, and an abstract that has been submitted to the ACM CHI Conference 
  on Human Factors in Computing Systems, Toronto, April 2014.}
}

\author{Lav~R.~Varshney, Florian Pinel, Kush~R.~Varshney, Debarun Bhattacharjya, Angela Sch\"{o}rgendorfer, and Yi-Min Chee%
\thanks{The authors are with the IBM Thomas J. Watson Research Center, Yorktown Heights, NY (e-mail: \{lrvarshn, pinel, krvarshn, debarunb, aschoer, ymchee\}@us.ibm.com).}}

\maketitle

\begin{abstract}
Computational creativity is an emerging branch of artificial intelligence that places computers 
in the center of the creative process.  Broadly, creativity involves a generative step to produce 
many ideas and a selective step to determine the ones that are the best.  Many previous attempts 
at computational creativity, however, have not been able to achieve a valid selective step.  This 
work shows how bringing data sources from the creative domain and from hedonic psychophysics 
together with big data analytics techniques can overcome this shortcoming to yield a system that 
can produce novel and high-quality creative artifacts.  Our data-driven approach is demonstrated 
through a computational creativity system for culinary recipes and menus we developed and deployed, 
which can operate either autonomously or semi-autonomously with human interaction.  We also comment 
on the volume, velocity, variety, and veracity of data in computational creativity.
\end{abstract}

\section{Introduction}
\IEEEPARstart{C}{reativity} is said to be the generation of a product or service that is judged to be novel and also to be 
appropriate, useful, or valuable by a knowledgeable social group \cite{Sawyer2012}, and 
is oft-said to be the pinnacle of intelligence.  Due to greater competitiveness in global markets 
for all industries, there is need to make product/service development cycles more efficient.  Indeed 
creativity is the basis for ``disruptive innovation and continuous re-invention'' \cite{IBM2010c}.  
Given limits on human creativity resources, it is important to develop technologies for  
greater creativity, either operating autonomously or in collaboration with people.

Can a computer be creative?  Computational creativity is concerned with 
machine systems that produce novel and high-quality artifacts (broadly construed) for 
the pleasure and consumption of people.  Such systems could produce jokes, poems, visual art, 
architectural blueprints, business processes, fashion ensembles, financial service designs, or any other 
such artifact that is popularly viewed by people as creative output.  In this paper, we focus 
primarily on culinary recipes, which include both the set and quantities of ingredients to be used 
as well as the methods and procedures of preparation.  We also discuss menus, which are sequences of 
culinary recipes.

We focus on a specific domain because at least with human creativity, there is substantial evidence that 
this cognitive ability is domain-specific \cite{KaufmanB2004}, in the sense a good engineer may not be 
a good poet.  Notwithstanding, psychometric testing has indicated some correlations of ability that allows 
domains to be grouped into categories: expressive creativity (visual arts, writing, humor); performance 
creativity (dance, drama, music); and scientific creativity (invention, science, culinary), with 
architecture not related to any of these \cite{CarsonPH2005}.  Most past attempts at computational 
creativity have focused on either expressive or performance creativity \cite{CardosoVW2009, Wiggins2006, 
Boden2010}, whereas we consider a form of scientific creativity 
(but see \cite{SchmidtL2009,KingROYABLMPSSWC2009}).

Our computational creativity system operates in stages that are modeled after stages in human creativity
\cite{Sawyer2012}: find problem, acquire knowledge, gather related information, incubate, generate ideas, 
combine ideas, select best ideas, and externalize ideas.  As may be apparent, many of these stages require 
the processing of external big data sets, as well as large volumes of intermediate data generated by the 
system itself.  The staged approach not only leads to modular system design, but also improves 
computer-human interaction when operating semi-autonomously.  Developing big data algorithms and systems 
is important, but interaction and presenting results to users in ways that allow them to trust insights 
is also important \cite{Fry2011}.  In semi-autonomous mode our system takes a \emph{mixed-initiative approach} 
where the human and computer have a creation conversation in which each contributes ideas \cite{SmithWM2011}, 
rather than the computer acting as a nanny, coach, or pen-pal for the human creator \cite{Lubart2005}.

For culinary creativity, we draw on a \emph{variety} of data sets: large repositories of existing recipes 
as inspiration, chemoinformatics data to understand food at the molecular level, and hedonic flavor 
psychophysics data to predict which compounds, ingredients, and dishes people will like and dislike.  
Since these data sets arise from noisy sensors like gas chromatography and from noisy data preprocessing
steps such as natural language processing, there may be issues of \emph{veracity} that algorithms
must be robust to, cf.~\cite{VarshneyVWM2013}.

These data sets are used to develop generative algorithms that intelligently produce thousands or millions
of new ideas from the recipe design space which, for particular dishes and regional cuisine influences, 
has a size in excess of $10^{24}$.   The large \emph{volume} of generated ideas must then be evaluated 
to select the best ones.  Evaluative metrics are based on principled models of human perception structured 
according to ideas from neurogastronomy and derived from recipe, chemical, and psychophysical data.  
The information-theoretic functional \emph{Bayesian surprise} is used to measure attraction of human 
attention and novelty.  Since the system is meant to support real-time interaction with human creators and
thereby make the product design cycle faster, the system must operate with the \emph{velocity} of thought.

Building on individual recipe design, complete menus can also be created using ideas from topic modeling.
By using the principle of variety across dishes in a menu, measured using a stochastic distance function,
input parameters for dish design may be generated and selected.

Many previous attempts at computational creativity had not been able to achieve a valid selective step 
\cite{Sawyer2012}, as we do here.  A central contribution of this work is in showing how bringing data 
sources from the creative domain and from hedonic psychophysics together with big data analytics techniques 
can overcome this shortcoming to yield a system that produces novel and high-quality creative artifacts.

It is worth noting that selection based on supervised learning trained on complete artifacts is not 
appropriate for creativity, as it is for search, since the entire premise is to create novel artifacts 
each time rather than to find existing ones.  Basic models of human perception, applied at the
constituent part level, and techniques for building up predictions of quality and novelty for whole 
artifacts is critical to our approach.

The remainder of the paper discusses details of data sets, data engineering, system architecture, 
data analytics algorithms, and results.  Although we use culinary recipes as the example domain herein, 
the basic concepts are generally applicable to big data approaches to computational creativity in
any domain.

\section{Creativity?}
\label{sec:creative:def}

Creativity seems to be a property of cognitive systems, 
but how should it be defined and assessed?  One aspect of cognition is understanding where in a museum a new painting 
hangs---reasoning about the world as it is, but another aspect is understanding whether that painting 
exhibits creativity---reasoning about things that had never previously been imagined.  Deductive and 
inductive reasoning about the world are easily assessed since there is often ground truth, but not so with creativity.

One definitional approach is to list several properties of a creative output, such as being novel, 
being useful, rejecting previously held ideas, and providing clarity \cite{NewellSS1963}.  But this definition does
not provide an operational method of assessment.  Viewing creativity as a relationship between the 
creator/creation and an observer \cite{Wiggins2006}, if a human evaluator deems something creative, we say 
it is creative \cite{PeaseC2011}.  Therefore, by definition, creativity is only meaningful in the presence 
of an audience perceiving the creation.  To formalize, we adopt a definition of creativity 
used in human creativity research.

\begin{defin}[\cite{Sawyer2012}]
\label{def:saw_create}
Creativity is the generation of a product that is judged to be novel and also to be appropriate, useful, 
or valuable by a suitably knowledgeable social group.
\end{defin}

In this definition, there are two dimensions: novelty and quality.  Although there is an individual component
to creativity, the adopted standard is a social one.  An artifact that is novel to the creator
need not be novel to the social group; what the creator finds of high-quality may not be seen as such
by the social group.  Thus creativity is fundamentally socially constructed.  A computational creativity 
system has no meaning in a closed universe devoid of people.  

The most common way to assess creativity of an artifact under this definition is the Consensual 
Assessment Technique (CAT) \cite{GetzelsC1976}, where the creativity of an artifact 
is rated by two or more experts in the field. The measured creativity
is the average rating of the judges.  Although it may seem this methodology is too subjective, 
many studies have demonstrated that ratings of experts are generally highly correlated, yielding good
interrater reliability \cite{KaufmanBC2009}.  In contrast, novice ratings are not highly correlated
and so novices should not be used for the CAT \cite{KaufmanBCS2008}.

An alternate definition for computational creativity would be by analogy to the Turing test---a system is 
creative if it produces artifacts indistinguishable from those produced by humans or having as much 
aesthetic value as those produced by humans \cite{Boden2010}.  We do not use this definition.

Our view is that a computational creativity machine without a way to evaluate its potential 
outputs is not really a computational creativity machine because generation and assessment must coexist 
for proper functioning.  In the same way that information cannot be encoded without a model of the receiver 
that will decode that information \cite{Shannon1950b}, artifacts cannot be created without a model of human evaluators.  

The lack of evaluative and selective ability has been a primary criticism of many previous computational
creativity systems.  Consider the computational creativity system for visual art
AARON.  AARON generates $150$ pieces a night, but H.~Cohen decides which $5$ to print by viewing 
them all: ``AARON doesn't choose its own criteria for what counts as a good painting\ldots 
To be considered truly creative, the program would have to develop its 
own selection criteria; Cohen was skeptical that this could ever happen'' \cite{Sawyer2012}. A computational 
creativity system for mathematical proofs AM suffers in the same way: ``The first and biggest problem 
is that AM generates a huge number of ideas, and most of them are boring or worthless; Lenat has to sort 
through all of the new ideas and select the ones that are good'' \cite{Sawyer2012}.

Fixing the operational view of Def.~\ref{def:saw_create}, we progress towards 
computational creativity system design \cite{Jordanous2012}.  As creativity is only meaningful in the presence of human 
perception, a human model is useful for the system to know whether or not it is 
producing creative artifacts and to guide its design process.  Such a component cannot be the final 
arbiter of creativity as that is a purely human determination, but it can be a very useful aid.  
A primary contribution of this work is to develop a data-driven evaluative/selective component.

\section{Stages of Human Creativity}
We use stages of human creativity to guide our computational creativity system design.
This will lead to a modular system architecture.  Note that stages are not always followed sequentially
by human creators; there can be backtracking and jumping around.

When the system operates in semi-autonomous mode, the computer acts as a colleague or partner to the human,
and so following the natural human process improves computer-human interaction.  Indeed 
there is an emerging consensus that even in purely human contexts, interacting groups are more creative than 
individuals, hence the value of computer-human interaction.

\begin{figure}
  \centering
  \includegraphics[width=3in]{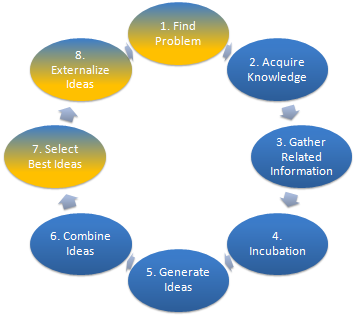}
  \caption{Stages in the cognitive process of creativity \cite{Sawyer2012}. In computational creativity, blue/gold stages may have 
	significant human-computer interaction whereas blue stages involve autonomous computer operation.}
  \label{fig:stages}
\end{figure}

We review stages of creativity delineated by Sawyer \cite{Sawyer2012}, given in Fig.~\ref{fig:stages} 
(which also depicts stages where human interaction is possible).

\begin{enumerate}
  \item \emph{Find the problem}: For ill-defined problems like creating new products, the first step 
is to actually identify and formulate the problem using \emph{divergent thinking}.  Exceptional creativity is
more likely when people work in areas where problems are not specified \emph{a priori}.

  \item \emph{Acquire knowledge}: The second stage is to 
learn everything there is to know about the problem, especially in terms of past creative artifacts.  
Without knowing what has already been done, there is no fodder for
inspiration nor is there a way to judge novelty.  Since it is impossible to be creative without first
internalizing the creative domain, data intake is necessary for creativity.

  \item \emph{Gather related information}: Besides learning about past examples of creative artifacts
within the domain, it is important to absorb information from a wide variety of other sources, so as to
link new information with existing problems and tasks.  

  \item \emph{Incubation}: In human creativity, it is important to give the mind the time to process 
	all of the gathered information, and to let the subconscious search for 
	new and appropriate combinations.

  \item \emph{Generate ideas}: After incubation, the mind is ready to generate ideas.  The generation 
of ideas is often considered the key step in creativity and is rather different from other forms of
reasoning such as induction or deduction.

  \item \emph{Combine ideas}: There is often value in cross-fertilization of ideas across problems 
and domains. Approaches to combining concepts across domains include attribute inheritance, property 
mapping, and concept specialization. 

  \item \emph{Select best ideas}: After a new idea or insight emerges, the creator must determine
whether it really is good.  This stage is sometimes referred to as 
\emph{convergent thinking}. In two-stage models, convergent thinking follows the divergent thinking 
phase of creativity.  The evaluation stage is fully conscious, drawing on large
amounts of domain knowledge to assess novelty and quality.

  \item \emph{Externalize the idea}: Successful creation requires not only ideas but also 
execution of those ideas, by identifying necessary resources to make them successful, forming plans 
for implementing the ideas, and so on.  This final stage is mostly conscious and directed. 
\end{enumerate}

This staged view of creativity forms the starting point for system architecture design.

\section{Creativity System Architecture}
In this section, we propose a system architecture for computational creativity that includes a 
data absorption and organization component, as well as a data-driven assessment component that 
models human perception.  

\begin{figure}
  \centering
  \includegraphics[width=3.5in]{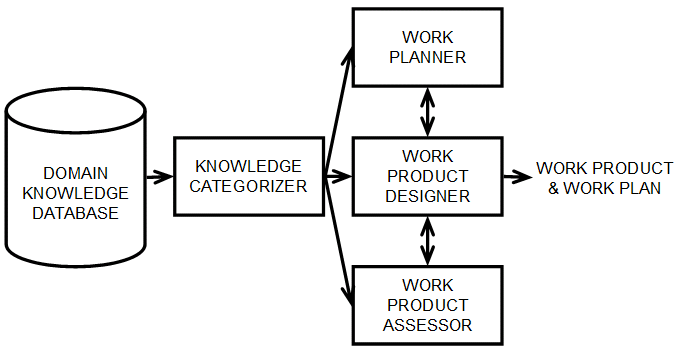}
  \caption{Block diagram of computational creativity system that produces a work product and a work plan.}
  \label{fig:blockdiagram}
\end{figure}

A block diagram for a proposed computational creativity system is presented in 
Fig.~\ref{fig:blockdiagram}, with three main data analytics components: a work planner, a work 
product designer, and a work product assessor, which interact to output a work product and work plan.  
These components are fed by a domain knowledge database and knowledge categorizer.
It is important to note that in our proposed system, the work planner and the work product assessor 
do not directly interact, but only do so through the work product designer.

The domain knowledge database represents information collected on the creative field of interest, 
including information on styles, tastes, constituents, combinations, evolution, 
regionality, culture, and methods of preparation.  It also includes a repository of existing 
artifacts that have been deemed creative by human audiences.  This knowledge is resolved and 
organized by the knowledge categorizer.  It is the source of data that the designer, planner, 
and assessor components draw from.  Information from related but distinct fields to the 
creative domain are also kept in the database.  As we will see, significant data engineering and natural
language processing is required for creating and using this knowledge database.

The designer generates new ideas for artifacts.  The assessor evaluates those potential design ideas 
for creativity and the planner determines the methods by which the ideas could be externalized.  All 
three components take input from the categorized database: the designer to draw inspiration for new 
ideas, the planner to learn from extant methods of preparation, and the assessor to evaluate a 
design idea against the repository of existing artifacts as well as against properties of 
constituents and combinations for creativity.

The designer is the lead component of the system.  Although it is possible to use human-like generative 
processes, a generation or design procedure wholly different from the human approach is valuable 
precisely because it would create things different from what a human would.  It may have 
different kinds of `illusions' or `blindspots' than a human, and thus would be a great supplement 
or support to human creativity.  These differences enlarge the hypothesis space and allow the machine 
to break new creative ground.

A creative computer is limited if it cannot evaluate 
proposed artifacts for creativity.  The assessor component models human perception, taste, and culture using data
analytics.  It examines creative ideas produced by the designer along two main dimensions: 
novelty and quality.  These metrics are defined on the basis of data sets within the creative domain, 
information related to the domain, and experimental data from hedonic psychophysics. 
It is worth noting that unlike many other data analytics problems and systems, computational creativity 
is fundamentally not a supervised learning problem.  One must decompose artifacts into parts and have 
assessment methods for the parts and for the recombination rules, to predict quality of completely new artifacts.  
There will not be training data available for novel complete artifacts.

Novelty can be assessed via information-theoretic or other similar quantifications of innovation 
within the context of all other existing artifacts in the domain of interest.  Quantifying quality 
requires a strong cognitive model because the quality of a creation truly is in the eye (or nose or tongue) 
of human beholders.  The novelty dimension is less specific to the particular creative domain of interest, 
whereas the quality dimension is intimately tied to it.  We provide details for a specific domain, the 
flavor of food, in the next section.

The final component, the work planner, determines steps needed to take concept to externalization.  
The work plan provides constraints on what designs are possible and can be optimized for efficient production,
e.g.\ using techniques from planning and operations research.  Generating the plan is itself a creative act 
and may be judged as such if an audience observes production.  However, artifacts can be deemed 
creative even if the work plan used to produce the artifact is not observed.

\section{Human Flavor Perception}
\label{sec:flavor}

Previous sections have been general; in this section we focus on the culinary domain and 
describe current understanding of human flavor perception, following the neurogastronomy
paradigm \cite{Shepherd2006}.

Human flavor perception is very complicated, involving a variety of external sensory stimuli and internal states 
\cite{Shepherd2006}.  Not only does it involve the five classical senses, but also sensing through the gut, and the 
emotional, memory-related, motivational, and linguistic aspects of food.  First of all there are the basic tastes: 
sweet, sour, salty, bitter, and umami.  The smell (both orthonasal and retronasal olfaction) of foods is the key contributor to 
flavor perception, which is in turn a property of the chemical compounds contained in the ingredients \cite{Burdock2009}.  
Olfactory perception is integrative rather than analytic, yielding unified percepts \cite{SnitzYWFKS2013}.  
There are typically tens to hundreds of different flavor compounds per food ingredient \cite{AhnABB2011}.  

Other contributors to flavor perception are the temperature, texture, astringency, 
and creaminess of the food; the color and shape of food; and the sound that the food makes.  The digestive 
system detects autonomic and metabolic properties of food.  Moreover, there are emotion, motivation, 
and craving circuits in the brain that influence flavor perception, which are in turn related to language, 
feeding, conscious flavor perception, and memory circuits.  Further, stimuli beyond the food itself, such 
as ambience of the room, influence flavor.

The complication in flavor perception is due to the interconnection and interplay between a multitude of neural 
systems, many of them not memoryless.  Recreating such a flavor perception system in a computer is an ambitious 
goal, but any progress is progress towards a viable computational creativity system 
for food.  Also, note that simply describing the factors and pathways of flavor perception fails 
to consider the settings of those factors that make food flavorful.  We return to this point in 
Sec.~\ref{sec:assessment}, where we propose methods for work product assessment motivated by human 
flavor perception. 

\section{Culinary Recipe Design}
\label{sec:recipe}

In the food domain, a dish is the basic unit of creation.  A cooked and plated dish is presented to a diner 
who perceives it and determines whether it is creative.  This presented dish can be the work product produced by 
a culinary computational creativity machine, as described generally in Fig.~\ref{fig:blockdiagram}.  The other 
output in Fig.~\ref{fig:blockdiagram}, the work plan, is a description of how to cook and how to plate the dish.  
A recipe is a work plan for how to cook a dish, but it is also a description of the work product, as it 
describes the ingredients to be used, their quantities, and their transformations and combinations.

A menu is a set of dishes that together constitute a meal. For example, a menu may consist of: asparagus soup, 
fillet of sole in lemon butter sauce, side of black beans with cilantro, cheesecake with coffee. 
A menu is an example of a sequence of artifacts where creativity is important; other examples of artifact sequences 
include albums of songs, comedy specials with several jokes, and sequences of clothing ensembles. 

Cutting-edge chefs must have impeccable culinary technique, but become renowned for their creative 
recipe designs.  A computer system to create novel and flavorful recipes as judged by people, would 
certainly be deemed creative.

The computer-generated culinary recipe design problem is not just one of locating existing recipes and recommending 
them \cite{TengLA2012}, but of creating new ones.  It is different from web search and product recommendation, and 
is truly part of an emerging computing paradigm distinct from fields such as information retrieval and statistical learning.  

Culinary computational creativity has recently been discussed in \cite{MorrisBBV2012}, where the authors focus 
only on soups rather than general recipes, and do not consider recipe assessment; in particular, they do not consider 
any of the neural, sensory, or psychological aspects of flavor.  In our recent previous work 
\cite{BhattacharjyaVPC2012}, we discuss general conceptions of novelty and flavor of dishes, but neither 
contextualize them nor present an overall system.

The overall culinary recipe design problem has many facets.  Through the lens of Fig.~\ref{fig:blockdiagram}, the 
first is to design and construct a suitable domain knowledge database.  This requires a data model enabling the 
system to reason about food and support algorithms for design, assessment, and planning.  In particular, it 
should be a repository of food ingredients and existing recipes, but also include knowledge about culinary 
styles and techniques, regional and seasonal cuisines, flavor compounds and their combinations, etc.  
We propose and discuss a data model for food in Section~\ref{sec:datamodel}.

A related aspect to building a computer chef is ingesting and processing raw data to populate the knowledge database 
structured according to the data model.  Sources include cookbooks and other repositories of recipes, culinary 
guides that explicate the culture of food, repositories of culinary techniques, and chemical databases of 
food ingredient constituents.  

Given a designed and populated domain knowledge database, a next step is developing a way to generate recipe ideas.  
Since cuisine naturally has evolutionary properties \cite{KinouchiDHZR2008}, i.e., cooking styles, techniques, 
and ingredient choices evolve and even exhibit features like the founder effect, genetic algorithms are 
one approach to the recipe design problem \cite{VeeramachaneniVO2012}.  Such an approach involves mutating 
and recombining existing recipes and can produce a myriad of potential recipes.

Besides random mutations and recombinations of recipes, there are some prominent culinary design principles 
that can be utilized.  For example, two principles focused on the chemosenses are the flavor pairing 
hypothesis \cite{AhnABB2011} and olfactory pleasantness maximization \cite{HaddadMRHS2010}.  Additional 
principles center around similarity of ingredients in properties such as geographic origin and seasonal origin.  
Chefs may also want to maintain balance (in terms of tastes, temperatures, or textures), or on the 
contrary accentuate a given characteristic, e.g.\ create the beefiest burger or the crunchiest cookie.

Finally, as discussed at the beginning of this section, a recipe is not only a work product but also a 
rudimentary work plan.  Therefore, in the culinary domain, a plan to produce the artifact is a must.  
The plan, utilizing a machine system's strengths, may be optimized and parallelized by formulating 
an operations research problem \cite{SwartD1981}.

\section{Data Engineering}

\subsection{Artifact Data Model}
\label{sec:datamodel}
Here we propose a data model that allows us to capture the salient pieces of domain knowledge 
to support all of the components of machine-generated creative recipe design.

\begin{figure*}
  \centering
  \includegraphics[width=7in]{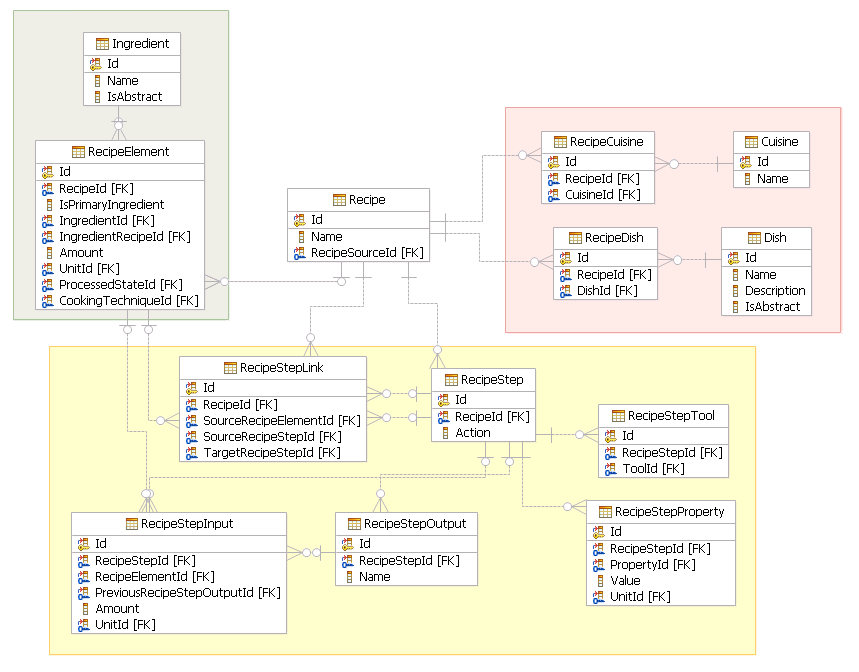}
  \caption{Knowledge representation schema for culinary recipes.  The ingredient component is expanded upon in Fig.~\ref{fig:ingredientschema}.}
  \label{fig:recipeschema}
\end{figure*}

\begin{figure*}
  \centering
  \includegraphics[width=7in]{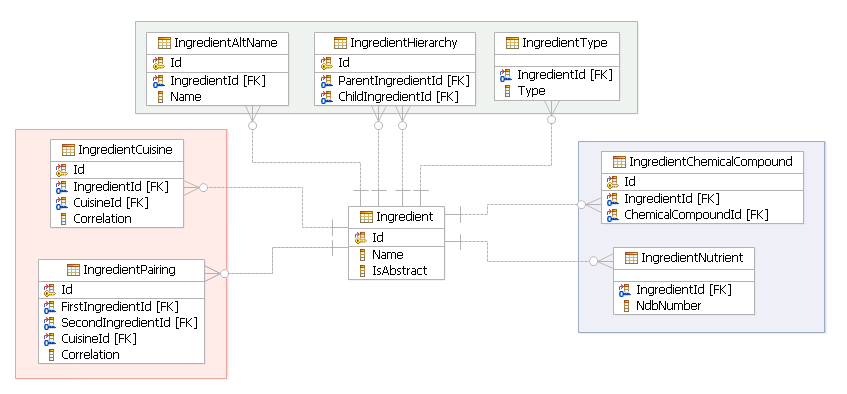}
  \caption{Knowledge representation schema for culinary ingredients that is a part of the overall schema for recipes given in Fig.~\ref{fig:recipeschema}.}
  \label{fig:ingredientschema}
\end{figure*}

As discussed in Sec.~\ref{sec:recipe}, the basic unit of cuisine is the dish, which is represented as 
a recipe.  We propose a representational model for culinary computational creativity that too has a 
recipe as the basic unit.  A schema---a codification of experience that includes a particular organized way 
of perceiving and responding to a complex situation---for cuisine that we propose is shown in 
Fig.~\ref{fig:recipeschema} and Fig.~\ref{fig:ingredientschema}.

Within this representation, we first capture the basic factors of the recipe, including the ingredients 
and their quantities, the tools required, and the sequence of cooking steps with input, output, tool and 
duration specified.  These basic factors are enough to be able to produce the artifact, i.e.~the dish.  
However, we need more elements in the representation to enable creative, flavorful idea generation by a computer.  

We must include knowledge about cultural context, human ratings, chemical analysis of ingredients and 
processes, and so on, to be able to characterize and emulate flavor perception.  For example, we include the 
name of the dish because it relates to the influence of cortical language circuits on flavor perception.  
We include the regional cuisine to which the dish belongs because regionality is a design principle in cooking.  
Similarly, we include the chemical flavor compound constituents of ingredients because flavor compound 
sharing is another design principle.  

As the preceding examples of data model elements illustrate, a creative culinary system's knowledge 
representation needs much more than simply a recounting of the ingredient list and cooking steps because 
it must reason about flavor perception, which involves many diverse sensing and memory pathways.  Idea 
generation can only use attributes in the data model and nothing more.  
It truly is the case that how the world is internally represented impacts what can be created.  
Creation, in our view, is the process of decomposing artifacts into their constituents as 
depicted in the data model, and then recomposing and reconstituting new artifact ideas.

Philosophically, schemata and diagrams define the universe within which cognition takes 
place \cite{Netz1999,Kaiser2005}.  Without a selective, simplified universe containing blindspots, 
the deployment of reasoning resources becomes untenable.  In the culinary 
case, we certainly have blindspots in our proposed schema.  For example, we do not include 
a data element about sound even though, as discussed in Sec.~\ref{sec:flavor}, 
it is a contributor to human flavor perception.  Since sound is not in the schema it is 
also outside the universe of reasoning for the system.  Importantly, it is purposefully not in the 
schema because capturing every component of flavor would be unmanageable.

\subsection{Natural Language Processing}

Recipes originally written in human readable format must be parsed to extract key knowledge for the data model \cite{TasseS2008}.
Beyond just ingredient lists as in other work \cite{TengLA2012,AhnABB2011}, our system needs 
to understand the inputs, outputs, tools, times, and techniques of the recipe steps.  

This is performed using natural language processing; our approach for processing ingredient amounts, names, and 
processed states is rule-based whereas our approach for processing the recipe instructions is based on statistical 
parsing with domain-specific tokens. Crowdsourcing has been used to develop an initial labeled corpus that can
be bootstrapped for improved statistical parsing.  As compared to training on general corpora (Wall Street Journal), 
naive statistical parsers trained on both general and domain-specific corpora can have accuracy that improves 
from $65$\% to $85$\%, in terms of getting the task, tool, ingredient, and tip correct from a recipe instruction sentence.

Recipes from data sets produced through peer production (as in the 25,000 recipes available on Wikia) 
come in various styles and are not as structured as recipes in published cookbooks, presenting extra challenges
we must handle.  Some notable attributes include personal commentary, multilingual text, missing information, 
abstracted description, and implied temporal information.

\subsection{Related Information}
Besides ingesting repositories of extant recipes, it is also important to gather related information.  One source
is Wikipedia, as a description of regional cuisines.  Again, natural language processing is needed
to convert text into insight, e.g.\ which ingredients are typical or canonical for a given region.  There
are hundreds of regional cuisines to be understood.

Another source of data, especially important for computational creativity, is hedonic psychophysics data
linked to chemical informatics data.  This provides characterizations of which flavor compounds are present
in which ingredients, and how much people like those flavors according to human psychophysics experiments.  
Each ingredient may contain hundreds of flavor compounds in varying concentrations, as determined in
the Volatile Compounds in Food 14.1 database (VCF) and in Fenaroli's Handbook of Flavor Ingredients 
as processed and released in \cite{AhnABB2011}.  

Since experimental psychophysics data may be sparse with respect to the thousands of
flavor compounds present in foods, data is also needed to predict the hedonic percepts of unmeasured
compounds.  This requires further physicochemical data on the various compounds; 
there can be hundreds or thousands of physicochemical descriptors such as the number of atoms
or the molecular complexity.  Much of this data is already structured in databases, but mapping named entities
across databases remains a problem that we solved manually.

\section{Data-Driven Assessment}
\label{sec:assessment}

We now turn to data-driven approaches for assessing novelty and flavor, which draw from human flavor perception 
science and operate within the universe set forth by the data model and related data.  We begin with a computational 
proposal for novelty, which can be applied more generally to other creative endeavors as well.  We then develop a 
computational quantification of pleasantness for food.  A creative recipe should have large values 
for novelty and pleasantness quantifications.

\subsection{Novelty}
\label{sec:assessment:novelty}

An artifact that is novel is surprising and changes the observer's world view.  Novelty can be quantified by 
considering a prior probability distribution of existing artifacts and the change in that probability distribution 
after the new artifact is observed, i.e.~the posterior probability distribution.  At the level of observable 
representation of artifacts, the difference between these probability distributions describes exactly how 
much the observer's world view has changed.  In recent work, such a quantitation has been given the name 
\emph{Bayesian surprise} and has been shown empirically to capture human notions of novelty and saliency 
across different modalities and levels of abstraction \cite{IttiB2009,BaldiI2010,Varshney2013}.

Surprise and novelty depend heavily on the observer's existing world view, and thus the same artifact may be 
novel to one observer and not novel to another observer.  That is why Bayesian surprise is measured as a change 
in the observer's specific prior probability distribution of known artifacts.  

Bayesian surprise is defined as follows.  Let $\mathcal{M}$ be the set of artifacts known to the observer, 
with each artifact in this repository being $M \in \mathcal{M}$.   Furthermore, a new artifact that is observed 
is denoted $A$.  The probability of an existing artifact is denoted $p(M)$, the conditional probability of the 
new artifact given the existing artifacts is $p(A|M)$, and via Bayes' theorem the conditional probability of 
the existing artifacts given the new artifact is $p(M|A)$.  The Bayesian surprise is defined as the 
following Kullback-Leibler divergence:
\begin{align}
\label{eq:Bs}
	\text{Bayesian surprise} &= D(p(M|A)\mid\mid p(M)) \\
		&= \int_{\mathcal{M}} p(M|A) \log\frac{p(M|A)}{p(M)}dM.\nonumber
\end{align}

Thinking of an artifact as an unordered tuple of $N$ ingredients, $A = \{I_1,\ldots,I_N\}$, combinatorial
expressions for probability distributions are found.  Although there are sophisticated techniques for estimating
information-theoretic functionals from data \cite{WangKV2009}, we find plug-in estimators to often be sufficient.
Handling the \emph{unseen elements} problem in statistical estimation \cite{OrlitskySZ2003}, however, is critical 
in computational creativity since the goal is to create completely novel artifacts.

\subsection{Flavor Pleasantness}
\label{sec:assessment:pleasantness}

The other dimension of creativity is the pleasantness of the flavors.  As noted in Sec.~\ref{sec:flavor}, 
knowledge of the senses and perceptual pathways gives insight into factors that determine the flavor of food.  
As noted there, constituent flavor compounds sensed by the olfactory system are the key to flavor perception.  
Thus a tractable step towards a data-driven model for flavor pleasantness is a model for odor pleasantness.  

Recent work has shown that there is a low-dimensional, almost scalar, hedonic quantity that describes the 
pleasantness of odors to humans, regardless of culture or other subjectivity \cite{KhanLFALHS2007,HaddadMRHS2010}.  
Moreover, this pleasantness is statistically associated with the physicochemical properties of compounds.  Hence 
we develop regression models to predict human-rated odor pleasantness of chemical compounds using their properties 
such as topological polar surface area, heavy atom count, complexity, rotatable bond count, and hydrogen bond acceptor count.  
Starting with tens of physicochemical features for $70$ observations in a pleasantness-labeled training data set \cite{HaddadMRHS2010},
multiple linear regression with model selection based on smallest prediction error in either 10-fold or leave-one-out 
cross validation yielded the small set of features used in the final regression model.

\begin{figure}
  \centering
  \includegraphics[width=3in]{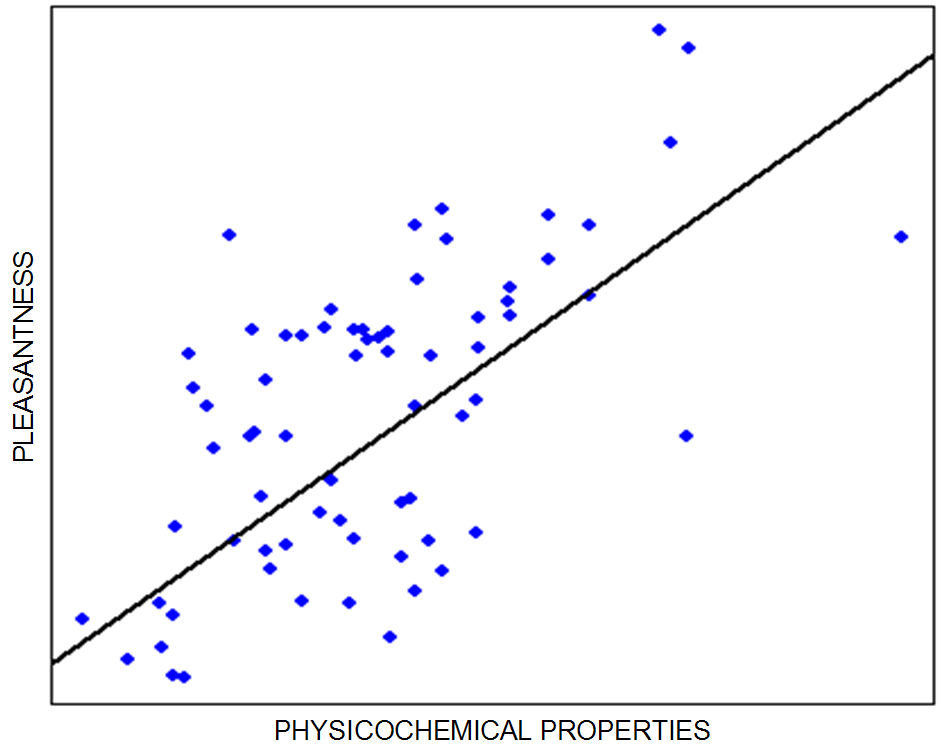}
  \caption{A regression model to predict pleasantness using the physicochemical properties of flavor compounds.}
  \label{fig:regression}
\end{figure}

The idea is shown in Fig.~\ref{fig:regression}, where data points are individual compounds, 
the vertical axis is the human-rated pleasantness, and the horizontal axis is a learned combination of 
chemical property features. Given a previously unrated compound, the regression model can be used to predict its pleasantness.

There is evidence that pleasantness is an approximately linear property of compounds \cite{LapidHS2008}.  If 
two compounds are mixed together and smelled, the hypothesis is that the odor pleasantness of the mixture is 
approximately a linear combination of the pleasantness values of the individual compounds.  With such linearity, 
one can predict the pleasantness of food ingredients that contain several flavor compounds and of dishes that 
in turn contain several ingredients.  The chemical properties of flavor compounds are well-catalogued and 
there is a growing body of literature cataloguing the flavor compound constituents of food ingredients \cite{AhnABB2011}.

Thus, if the recipe assessor is given a proposed idea by the recipe designer in a computational creativity system, 
it can calculate its novelty using Bayesian surprise and calculate its flavorfulness using an olfactory pleasantness 
regression model applied to its constituent ingredients and flavor compounds in those ingredients.  
Such scoring represents a data-driven approach to assessing artifacts that have been newly created and have never 
existed before.  

\section{Computer-Human Interaction for Semi-autonomy}
Although the computational creativity system defined thus far can operate autonomously, it can have greater
impact as part of an integrated collaborative work flow with human creators.  We implement an interactive interface,
taking a mixed-initiative approach to human-computer interaction via turns between human and computer \cite{SmithWM2011}.

\begin{figure}
  \centering
  \includegraphics[width=3.8in]{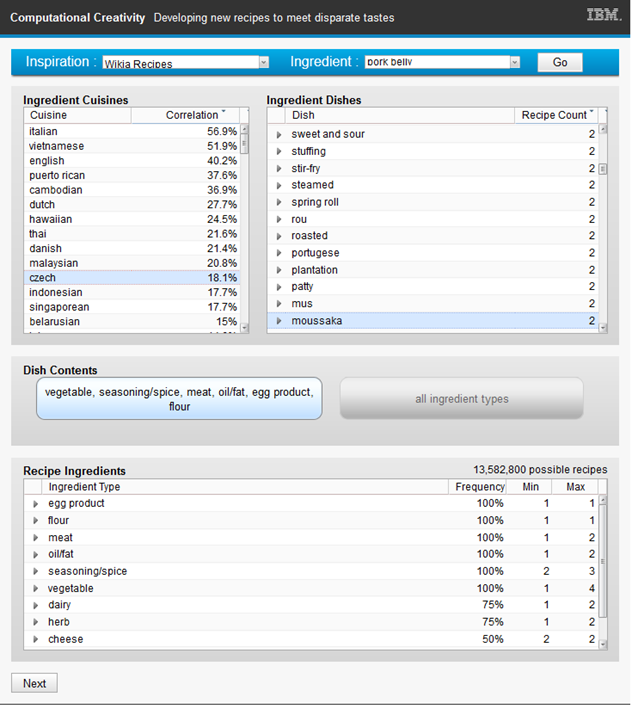}
  \caption{Interface for problem-finding, showing common and uncommon choices.}
  \label{fig:problemfinding}
\end{figure}

The first step in creativity is problem-finding.  Mediated by a novel interactive interface design, this may be accomplished 
jointly by the human and the machine, by picking a key ingredient, one or more regional cuisines to influence flavor, 
and a dish type such as soup or quiche.  Machine learning is used to suggest ingredient types, though this can be modified 
by the human.  The problem-finding input screen, Fig.~\ref{fig:problemfinding}, sets parameters for the generative algorithm 
to create thousands or millions of ingredient list ideas.

\begin{figure}
  \centering
  \includegraphics[width=3.8in]{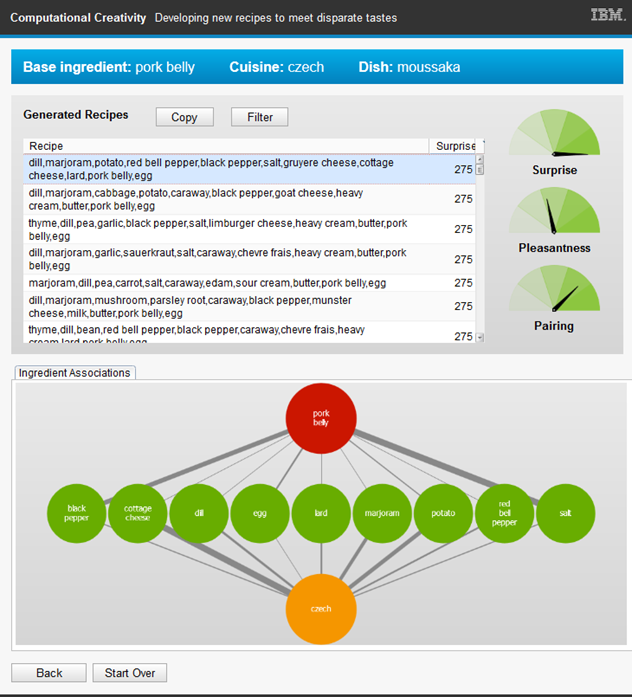}
  \caption{Interface for selection, showing design reasoning and ratings along dimensions of novelty and quality.}
  \label{fig:selection}
\end{figure}

The penultimate stage in creativity is selecting the best idea(s).  The computer predicts which generated recipes 
will be the most surprising to a human observer, will be perceived as the most flavorful, and will have the best 
pairings of ingredients (see \cite{AhnABB2011}).  These metrics are used to rank the generated ideas and then a 
human makes the final selection, see the selection screen: Fig.~\ref{fig:selection}.  In our experience, humans 
select one of the top ten ideas, rather than looking through hundreds or thousands of possibilities.  
Hence selection is truly a collaboration between human and machine.

Visualizations at the bottom of the screen help the human understand the reasoning used by the computer in generating
and ranking ideas, so as to provide confidence.  This includes visualizing the design process, as well as 
the metrics of pleasantness, pairing, and surprise.

\begin{figure}
  \centering
  \includegraphics[width=3.8in]{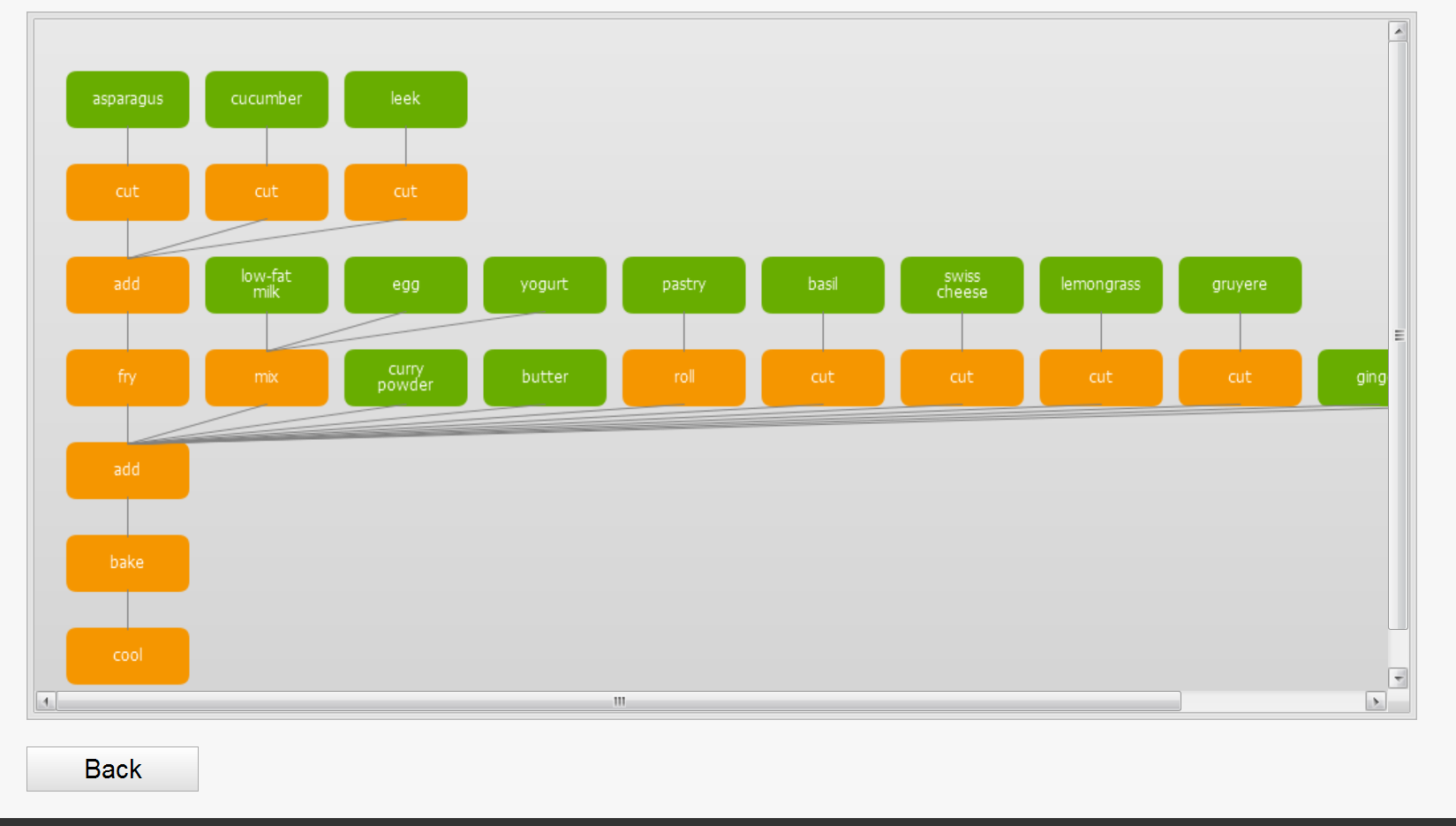}
  \caption{Bottom of interface for externalization, showing recipe steps and their partial ordering.  Green boxes 
are recipe ingredients and orange boxes represent actions performed. For the example quiche recipe depicted: 
vegetables are cut and fried together, wet ingredients are mixed, pie crust dough is rolled, etc. 
Steps can be performed by multiple cooks in parallel, until all elements are put together, baked, and cooled.}
  \label{fig:steps}
\end{figure}

The final stage of creativity is externalizing.  In recipe creation, this involves coming up with not just 
the list of ingredients (the focus of idea generation and selection), but also proportions and 
recipe steps.  Professional chefs often operate without computer support for externalizing, but amateurs
appreciate guidance since it too requires significant creativity.  The final screen shows proportions 
and steps in the form of a directed acyclic graph, Fig.~\ref{fig:steps}.  
Possible actions are abstracted to improve reasoning.

\section{Menus of Recipes}
So far we have discussed creating a single recipe at a time, and in the previous section problem finding was cast
as human-machine interaction for picking a key ingredient, regional cuisines, and dish types.  When creating a sequence
of dishes, such parameters should be linked across dishes.  Here we introduce the notion of dish \emph{variety}, 
which we consider an important aspect of creative menus.  Taking a big data approach,
we use a modeling technique based on topic modeling. Topic models are used to identify underlying latent topics 
in a set of documents; we apply them to a repository of recipes.

\subsection{Topic Modeling}
Def.~\ref{def:saw_create} requires creative artifacts to be novel and of high quality. For menus, the novelty and quality 
of the set is partially determined by its constituent dish recipes, but variety is a property of multiple artifacts: 
it is an emergent property for collections and is not definable for individual artifacts.  

Topic models are machine learning algorithms that discover the main underlying themes that pervade 
a large collection of documents through generative model assuming documents are probabilistic mixtures of a set of 
underlying latent variables, i.e.\ ``topics'', and the ``words'' that comprise a document are probabilistically 
generated from these topics \cite{Blei2012}. Here we treat recipes as documents, and apply the Latent Dirichlet 
Allocation (LDA) method of topic modeling \cite{BleiNJ2003} to the Wikia corpus of recipes to indicate how big data 
approaches can be used for problem finding.  While applying this method, we assume that a recipe is adequately summarized 
as a set of ingredients, but see Sec.~\ref{sec:datamodel}. Topic modeling has previously been applied to recipes  
for other purposes \cite{Nedovic2013}.

\begin{figure}
  \centering
  \includegraphics[width=3.5in]{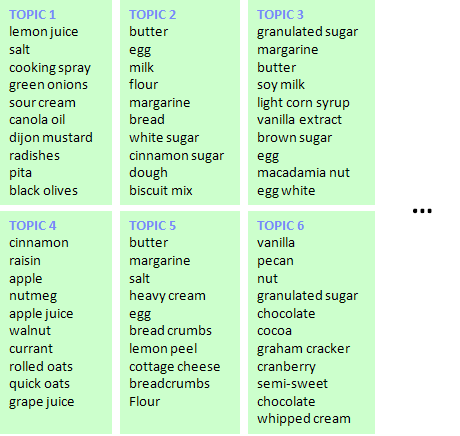}
  \caption{Example topics when the LDA method is applied to Wikia recipes.}
  \label{fig:topics}
\end{figure}

Learned recipe topics can be interpreted as some ingredient combinations that either go well together (like sauces) 
or that can be substituted (like a set of fruits). To generate a new recipe, one selects an ingredient by first choosing 
an underlying recipe topic, and then drawing the ingredient from the recipe topic-specific distribution. 
Fig.~\ref{fig:topics} shows a list of a select few topics that were generated using the LDA method applied 
to the Wikia corpus, along with some of the most likely ingredients in these topics. 

\subsection{Assessing Variety}
Here we propose a topic-based approach for assessing variety in menus. Consider a menu with $K$ recipes, where the $k$th recipe 
is $A_k = \{I_{k_1},\ldots,I_{k_N}\}$ and $I_{k_n}$ is the $n$th ingredient in the $k$th recipe.  For notational convenience, assume 
all recipes in the menu have the same number of ingredients $N$, but the method applies to the general case where recipes have varying 
numbers of ingredients.  Suppose that a topic model with $L$ underlying recipe topics has been used to model the generative process 
by which the corpus of recipes was created. Let $T$ denote the random variable for the marginal distribution of the recipe topics 
from which an ingredient is selected for a recipe, and let $t_{\ell}$ be the $\ell$th topic. 

Note that a topic model is a Bayesian model that considers the relationship between the parameters, the recipe topics and the ingredients 
that are chosen in recipes. Therefore, we can use Bayes' rule and perform inference to compute the probability that a particular recipe 
topic was selected in picking a particular ingredient. Let $P(T|I_{k_n})$ denote the probability distribution over recipe topics for the
$n$th ingredient in the $k$th recipe. This probability measures how an ingredient is associated with the underlying themes in the recipe 
database. 

To measure variety in menus, we compare how various recipe topics are spanned by the constituent recipes through the notion of a 
\emph{topic spanning metric} for a recipe, which measures how that particular recipe is associated with the various recipe topics. 
This topic spanning metric for a recipe can be any function of topic probabilities for its constituent ingredients: 
\begin{equation}
s_k = s(A_k) = f\left[P(T|I_{k_1}),\ldots,P(T|I_{k_N}) \right] \mbox{.}
\end{equation} 
An example is when a topic is said to be covered by a recipe when at least $1$ ingredient in that recipe was selected from that 
topic. Let $s_{k_{\ell}}$ be the probability that the $\ell$th topic was used by at least $1$ ingredient in the $k$th recipe, in which case:
\begin{equation}
s_{k_{\ell}} = 1 - \prod_{n=1}^N \left[1 - P(T = t_{\ell} |I_{k_n}) \right] \mbox{.}
\end{equation} 

The following vector is a potential topic spanning metric; it measures the degree to which every recipe topic is associated with the 
$k$th recipe: $s_k = \{s_{k_{\ell}}\}_{\ell = 1}^L$. We can now score menu variety based on the distance between spanning metrics for 
recipes in the menu. Note that all recipes have vectors of the same dimension, which is the number of recipes topics.  
\begin{equation}
\label{eq:var}
\text{Variety} = D[s_1,\ldots,s_K]\mbox{,}
\end{equation}
where $D[\cdot]$ is a distance metric like Euclidean distance.  The variety score computed in this fashion assesses how 
recipes in the menu differ from each other in terms of the fundamental underlying themes from which they were generated. 
Thus the topic modeling approach allows for the effective use of data to identify and model variety in menus.

\section{Variety, Veracity, Volume, and Velocity}

It has been said that ``Big Data is all about better analytics on a broader spectrum of data'' \cite{ZikopoulosDPDCG2013}
so as to provide insight and value.  In particular, the four ``V''s of Big Data: variety, veracity, volume, and velocity
must all be considered to derive maximal value in data-driven approaches that push computing to the emerging area of 
cognition and creativity \cite{KellyH2013}.

Variety is about trying to capture all of the data that pertains to the cognitive process within the creative domain and 
also outside of it.  Since most data is semistructured or unstructured, making sense of it is not natural to computers and
requires new ways of data engineering and data management.  What is required is a holistic approach that allows
the ability to mash together different kinds of data and act on them to create artifacts which have never been imagined.

Veracity refers to quality or trustworthiness of data.  Since much of the data we have access to is not complete, precise, 
or certain, see e.g.\ \cite{VarshneyVWM2013}, we must embrace analytics algorithms that are robust so as to allow drawing 
the right conclusions.

The sheer volume of intermediate data that is generated as part of computational creativity is a challenge in itself.
Moreover, operating at the velocity of human thought is crucial for a computational creativity system to serve as a
true colleague to human creators and to quicken the product design cycle.  Efficient computer architectures and paralellizable 
algorithms can help mitigate both of these concerns.  

\section{Summary, Results, and Outlook}
\noindent{\bf Summary} In this paper, we have developed a computational creativity system that can automatically or semi-automatically design and 
discover culinary recipes that are flavorful, novel, and perhaps healthy.  This is done through artificial intelligence
algorithms based on Bayesian probability and regression analysis that draw on big data techniques, as well as disparate data sources 
from culinary traditions, chemoinformatics, and hedonic psychophysics.  We proposed a structure for a computational creativity system 
that contains three main components: a designer, an assessor and a planner, all fed by a domain knowledge database.  
Furthermore, we discussed the role of the domain knowledge database in structuring and setting the bounds for cognitive processing.
\vspace{2mm}

\noindent{\bf Results} Recipes created by the computational creativity system, such as a Caymanian Plantain Dessert, have been rated as more creative
than existing recipes in online repositories using a method similar to the Consensual Assessment Technique \cite{Davis2013}.
Moreover, professional chefs at various hotels, restaurants, and culinary schools have indicated that the system helps them
explore new vistas in food.  These results provide validation for the data-driven approach to computational creativity.
\vspace{2mm}

\noindent{\bf Outlook} Creativity is easy neither for people nor for machines, but the challenges are different.  Without taking advantage of 
modularity, people often have trouble being creative and innovative because they are overwhelmed by the combinatorial 
complexity of large design spaces \cite{McNerneyFRT2011}.  Since people end up thinking modularly, progression of creative 
thought is often evolutionary \cite{Basalla1988}.  A computational creativity system can test quadrillions of ideas at once 
without needing to invoke modularity and may thus offer solutions that completely redefine an art.  Such creations may offer 
advantages by being completely `outside the box' through large jumps in thought rather than gradual evolutionary changes.  

Although we took a particular creative application domain---culinary recipe design---as an example, the system architectures, approaches,
and insights garnered in facing the challenges should be applicable across creative domains.  

\section*{Acknowledgment}
The authors thank Lamia Tounsi, Krishna Ratakonda, Jun Wang, EE Jan, and Terry Heath from IBM, as well as 
James Briscione and Michael Laiskonis from the Institute of Culinary Education.

\bibliographystyle{IEEEtran} 
\bibliography{abrv,conf_abrv,lrv_lib}

\begin{thebibliography}{10}
\providecommand{\url}[1]{#1}
\csname url@samestyle\endcsname
\providecommand{\newblock}{\relax}
\providecommand{\bibinfo}[2]{#2}
\providecommand{\BIBentrySTDinterwordspacing}{\spaceskip=0pt\relax}
\providecommand{\BIBentryALTinterwordstretchfactor}{4}
\providecommand{\BIBentryALTinterwordspacing}{\spaceskip=\fontdimen2\font plus
\BIBentryALTinterwordstretchfactor\fontdimen3\font minus
  \fontdimen4\font\relax}
\providecommand{\BIBforeignlanguage}[2]{{%
\expandafter\ifx\csname l@#1\endcsname\relax
\typeout{** WARNING: IEEEtran.bst: No hyphenation pattern has been}%
\typeout{** loaded for the language `#1'. Using the pattern for}%
\typeout{** the default language instead.}%
\else
\language=\csname l@#1\endcsname
\fi
#2}}
\providecommand{\BIBdecl}{\relax}
\BIBdecl

\bibitem{VarshneyPVSC2013}
L.~R. Varshney, F.~Pinel, K.~R. Varshney, A.~Sch{\"o}rgendorfer, and Y.-M.
  Chee, ``Cognition as a part of computational creativity,'' in \emph{Proc.
  12th IEEE Int. Conf. Cogn. Inform. Cogn. Comput. (ICCI*CC 2013)}, Jul. 2013,
  pp. 36--43.

\bibitem{Sawyer2012}
R.~K. Sawyer, \emph{Explaining Creativity: The Science of Human
  Innovation}.\hskip 1em plus 0.5em minus 0.4em\relax Oxford: Oxford University
  Press, 2012.

\bibitem{IBM2010c}
{IBM}, ``Capitalizing on complexity: Insights from the global chief executive
  officer study,'' May 2010.

\bibitem{KaufmanB2004}
J.~C. Kaufman and J.~Baer, \emph{Creativity Across Domains: Faces of the
  Muse}.\hskip 1em plus 0.5em minus 0.4em\relax Psychology Press, 2004.

\bibitem{CarsonPH2005}
S.~H. Carson, J.~B. Peterson, and D.~M. Higgins, ``Reliability, validity, and
  factor structure of the creative achievement questionnaire,''
  \emph{Creativity Res. J.}, vol.~17, no.~1, pp. 37--50, 2005.

\bibitem{CardosoVW2009}
A.~Cardoso, T.~Veale, and G.~A. Wiggins, ``Converging on the divergent: The
  history (and future) of the {I}nternational {J}oint {W}orkshops in
  {C}omputational {C}reativity,'' \emph{A. I. Mag.}, vol.~30, no.~3, pp.
  15--22, Fall 2009.

\bibitem{Wiggins2006}
G.~A. Wiggins, ``Searching for computational creativity,'' \emph{New Generat.
  Comput.}, vol.~24, no.~3, pp. 209--222, Sep. 2006.

\bibitem{Boden2010}
M.~A. Boden, ``The {T}uring test and artistic creativity,'' \emph{Kybernetes},
  vol.~39, no.~3, pp. 409--413, 2010.

\bibitem{SchmidtL2009}
M.~Schmidt and H.~Lipson, ``Distilling free-form natural laws from experimental
  data,'' \emph{Science}, vol. 324, no. 5923, pp. 81--85, Apr. 2009.

\bibitem{KingROYABLMPSSWC2009}
R.~D. King, J.~Rowland, S.~G. Oliver, M.~Young, W.~Aubrey, E.~Byrne,
  M.~Liakata, M.~Markham, P.~Pir, L.~N. Soldatova, A.~Sparkes, K.~E. Whelan,
  and A.~Clare, ``The automation of science,'' \emph{Science}, vol. 324, no.
  5923, pp. 85--89, Apr. 2009.

\bibitem{Fry2011}
C.~Fry, ``Closing the gap between analytics and action,'' \emph{INFORMS
  Analytics Mag.}, vol.~4, no.~6, pp. 4--5, Nov.-Dec. 2011.

\bibitem{SmithWM2011}
G.~Smith, J.~Whitehead, and M.~Mateas, ``Computers as design collaborators:
  Interacting with mixed-initiative tools,'' in \emph{Proc. ACM Creat. Cogn.
  Worshop Semi-Autom. Creat. (SAC 2011)}, Nov. 2011.

\bibitem{Lubart2005}
T.~Lubart, ``How can computers be partners in the creative process,''
  \emph{Int. J. Hum.-Comput. Studies}, vol.~63, no. 4--5, pp. 365--369, Oct.
  2005.

\bibitem{VarshneyVWM2013}
K.~R. Varshney, L.~R. Varshney, J.~Wang, and D.~Meyers, ``Flavor pairing in
  {M}edieval {E}uropean cuisine: A study in cooking with dirty data,'' in
  \emph{Proc. Int. Joint Conf. Artif. Intell. Workshops}, Aug. 2013, pp. 3--12.

\bibitem{NewellSS1963}
A.~Newell, J.~C. Shaw, and H.~A. Simon, ``The process of creative thinking,''
  in \emph{Contemporary Approaches to Creative Thinking}, H.~E. Gruber,
  G.~Terrell, and M.~Wertheimer, Eds.\hskip 1em plus 0.5em minus 0.4em\relax
  New York: Atherton, 1963, pp. 63--119.

\bibitem{PeaseC2011}
A.~Pease and S.~Colton, ``On impact and evaluation in computational creativity:
  A discussion of the {T}uring test and an alternative proposal,'' in
  \emph{Proc. Symp. Comput. Phil., AISB'11 Conv.}, Apr. 2011.

\bibitem{GetzelsC1976}
J.~W. Getzels and M.~Csikszentmihalyi, \emph{The Creative Vision: Longitudinal
  Study of Problem Finding in Art}.\hskip 1em plus 0.5em minus 0.4em\relax New
  York: John Wiley \& Sons, 1976.

\bibitem{KaufmanBC2009}
J.~C. Kaufman, J.~Baer, and J.~C. Cole, ``Expertise, domains, and the
  consensual assessment technique,'' \emph{J. Creative Behav.}, vol.~43, no.~4,
  pp. 223--233, Dec. 2009.

\bibitem{KaufmanBCS2008}
J.~C. Kaufman, J.~Baer, J.~C. Cole, and J.~D. Sexton, ``A comparison of expert
  and nonexpert raters using the consensual assessment technique,''
  \emph{Creativity Res. J.}, vol.~20, no.~2, pp. 171--178, 2008.

\bibitem{Shannon1950b}
C.~E. Shannon, ``The redundancy of {E}nglish,'' in \emph{Trans. 7th Conf.
  Cybern.}, Mar. 1950, pp. 123--158.

\bibitem{Jordanous2012}
A.~K. Jordanous, ``Evaluating computational creativity: A standardised
  procedure for evaluating creative systems and its application,'' Ph.D.
  dissertation, University of Sussex, Dec. 2012.

\bibitem{Shepherd2006}
G.~M. Shepherd, ``Smell images and the flavour system in the human brain,''
  \emph{Nature}, vol. 444, no. 7117, pp. 316--321, Nov. 2006.

\bibitem{Burdock2009}
G.~A. Burdock, \emph{Fenaroli's Handbook of Flavor Ingredients}.\hskip 1em plus
  0.5em minus 0.4em\relax Boca Raton, FL: CRC Press, 2009.

\bibitem{SnitzYWFKS2013}
K.~Snitz, A.~Yablonka, T.~Weiss, I.~Frumin, R.~M. Khan, and N.~Sobel,
  ``Predicting odor perceptual similarity from odor structure,'' \emph{{PLoS}
  Comput. Biol.}, vol.~9, no.~9, p. e1003184, Sep. 2013.

\bibitem{AhnABB2011}
Y.-Y. Ahn, S.~E. Ahnert, J.~P. Bagrow, and A.-L. Barab\'{a}si, ``Flavor network
  and the principles of food pairing,'' \emph{Sci. Reports}, vol.~1, p. 196,
  Dec. 2011.

\bibitem{TengLA2012}
C.-Y. Teng, Y.-R. Lin, and L.~A. Adamic, ``Recipe recommendation using
  ingredient networks,'' in \emph{Proc. 3rd Annu. ACM Web Sci. Conf.
  (WebSci'12)}, Jun. 2012, pp. 298--307.

\bibitem{MorrisBBV2012}
R.~G. Morris, S.~H. Burton, P.~M. Bodily, and D.~Ventura, ``Soup over beans of
  pure joy: Culinary ruminations of an artificial chef,'' in \emph{Proc. Int.
  Conf. Comput. Creativity (ICCC 2012)}, May 2012, pp. 119--125.

\bibitem{BhattacharjyaVPC2012}
D.~Bhattacharjya, L.~R. Varshney, F.~Pinel, and Y.-M. Chee, ``Computational
  creativity: A two-attribute search technique,'' in \emph{INFORMS Annu.
  Meeting}, Oct. 2012.

\bibitem{KinouchiDHZR2008}
O.~Kinouchi, R.~W. Diez-Garcia, A.~J. Holanda, P.~Zambianchi, and A.~C. Roque,
  ``The non-equilibrium nature of culinary evolution,'' \emph{New J. Phys.},
  vol.~10, p. 073020, 2008.

\bibitem{VeeramachaneniVO2012}
K.~Veeramachaneni, E.~Vladislavleva, and U.-M. {O'}Reilly, ``Knowledge mining
  sensory evaluation data: genetic programming, statistical techniques, and
  swarm optimization,'' \emph{Genet. Program. Evolvable Mach.}, vol.~13, no.~1,
  pp. 103--133, Mar. 2012.

\bibitem{HaddadMRHS2010}
R.~Haddad, A.~Medhanie, Y.~Roth, D.~Harel, and N.~Sobel, ``Predicting odor
  pleasantness with an electronic nose,'' \emph{{PLoS} Comput. Biol.}, vol.~6,
  no.~4, p. e1000740, Apr. 2010.

\bibitem{SwartD1981}
W.~Swart and L.~Donno, ``Simulation modeling improves operations, planning, and
  productivity of fast food restaurants,'' \emph{Interfaces}, vol.~11, no.~6,
  pp. 35--47, Dec. 1981.

\bibitem{Netz1999}
R.~Netz, \emph{The Shaping of Deduction in Greek Mathematics: A Study in
  Cognitive History}.\hskip 1em plus 0.5em minus 0.4em\relax Cambridge:
  Cambridge University Press, 1999.

\bibitem{Kaiser2005}
D.~Kaiser, \emph{Drawing Theories Apart: The Dispersion of Feynman Diagrams in
  Postwar Physics}.\hskip 1em plus 0.5em minus 0.4em\relax Chicago: University
  of Chicago Press, 2005.

\bibitem{TasseS2008}
D.~Tasse and N.~A. Smith, ``{SOUR CREAM}: Toward semantic processing of
  recipes,'' Carnegie Mellon University, Pittsburgh, Tech. Rep. CMU-LTI-08-005,
  May 2008.

\bibitem{IttiB2009}
L.~Itti and P.~Baldi, ``{B}ayesian surprise attracts human attention,''
  \emph{Vis. Res.}, vol.~49, no.~10, pp. 1295--1306, Jun. 2009.

\bibitem{BaldiI2010}
P.~Baldi and L.~Itti, ``Of bits and wows: A {B}ayesian theory of surprise with
  applications to attention,'' \emph{Neural Netw.}, vol.~23, no.~5, pp.
  649--666, Jun. 2010.

\bibitem{Varshney2013}
L.~R. Varshney, ``To surprise and inform,'' in \emph{Proc. 2013 IEEE Int. Symp.
  Inf. Theory}, Jul. 2013, pp. 3145--3149.

\bibitem{WangKV2009}
Q.~Wang, S.~R. Kulkarni, and S.~{Verd\'{u}}, ``Universal estimation of
  information measures for analog sources,'' \emph{Found. Trends Commun. Inf.
  Theory}, vol.~5, no.~3, pp. 265--353, 2009.

\bibitem{OrlitskySZ2003}
A.~Orlitsky, N.~P. Santhanam, and J.~Zhang, ``Always {G}ood {T}uring:
  Asymptotically optimal probability estimation,'' \emph{Science}, vol. 302,
  no. 5644, pp. 427--431, Oct. 2003.

\bibitem{KhanLFALHS2007}
R.~M. Khan, C.-H. Luk, A.~Flinker, A.~Aggarwal, H.~Lapid, R.~Haddad, and
  N.~Sobel, ``Predicting odor pleasantness from odorant structure: Pleasantness
  as a reflection of the physical world,'' \emph{J. Neurosci.}, vol.~27,
  no.~37, pp. 10\,015--10\,023, Sep. 2007.

\bibitem{LapidHS2008}
H.~Lapid, D.~Harel, and N.~Sobel, ``Prediction models for the pleasantness of
  binary mixtures in olfaction,'' \emph{Chem. Senses}, vol.~33, no.~7, pp.
  599--609, Sep. 2008.

\bibitem{Blei2012}
D.~M. Blei, ``Probabilistic topic modeling,'' \emph{Commun. ACM}, vol.~55,
  no.~4, pp. 77--84, Apr. 2012.

\bibitem{BleiNJ2003}
D.~M. Blei, A.~Y. Ng, and M.~I. Jordan, ``Latent {D}irichlet allocation,''
  \emph{J. Mach. Learn. Res.}, vol.~3, pp. 993--1022, Mar. 2003.

\bibitem{Nedovic2013}
V.~Nedovic, ``Learning recipe ingredient space using generative probabilistic
  models,'' in \emph{Proc. Int. Joint Conf. Artif. Intell. Workshops}, Aug.
  2013, pp. 13--18.

\bibitem{ZikopoulosDPDCG2013}
P.~Zikopoulos, D.~Deroos, K.~Parasuraman, T.~Deutsch, D.~Corrigan, and
  J.~Giles, \emph{Harness the Power of Big Data}.\hskip 1em plus 0.5em minus
  0.4em\relax New York: McGraw Hill, 2013.

\bibitem{KellyH2013}
J.~E. Kelly, III and S.~Hamm, \emph{Smart Machines: {IBM}'s {W}atson and the
  Era of Cognitive Computing}.\hskip 1em plus 0.5em minus 0.4em\relax New York:
  Columbia University Press, 2013.

\bibitem{Davis2013}
A.~P. Davis, ``Digital gastronomy: When an {IBM} algorithm cooks, things get
  complicated--and tasty,'' \emph{Wired}, vol.~20, no.~10, Oct. 2013.

\bibitem{McNerneyFRT2011}
J.~McNerney, J.~D. Farmer, S.~Redner, and J.~E. Trancik, ``Role of design
  complexity in technology improvement,'' \emph{Proc. Natl. Acad. Sci. U.S.A.},
  vol. 108, no.~22, pp. 9008--9013, May 2011.

\bibitem{Basalla1988}
G.~Basalla, \emph{The Evolution of Technology}.\hskip 1em plus 0.5em minus
  0.4em\relax New York: Cambridge University Press, 1988.

\end{thebibliography}

\end{document}